# Adaptive Load-Aware Sampling for Network Monitoring on Multicore Commodity Hardware

Lothar Braun, Cornelius Diekmann, Nils Kammenhuber, Georg Carle
{braun,diekmann,kammenhuber,carle}@net.in.tum.de
Technische Universität München

*Abstract*—Many current traffic monitoring systems employ deep packet inspection (DPI) in order to analyze network traffic. These systems include intrusion detection systems, software for network traffic accounting, traffic classification, or systems for monitoring service-level agreements. Traffic volumes and link speeds of current enterprise and ISP networks transform the process of inspecting traffic payload into a challenging task. In this paper we propose a novel adaptive sampling algorithm that selects the maximum number of packets from the network that the DPI system is able to consume. Our algorithm adapts its sampling rate according to the network traffic currently observed, and the number of packets that a monitoring application is able to process. It can be used in conjunction with current multicore-aware network traffic analysis setups, which allow for exploiting current multi-core hardware. We show the applicability of our algorithm with live-tests on a heavily used 10G link with real network monitoring tools.

## I. INTRODUCTION

Deep Packet Inspection (DPI) is a core component in numerous state-of-the-art approaches for network monitoring. Well-known intrusion detection systems such as Snort [1] or Bro [2] rely on inspecting packet payload for detecting attack traffic or policy violations. Recent developments that focus on detecting botnets employ payload inspection as well: Gu et al., for example, use Snort to generate botnet-specific alerts and propose a correlation engine that tries to find botnet-like behavior in the detected events [3]. Other techniques scan for specific botnet command and control channels [4], or implement generic payload inspection algorithms to find groups of similar behaving hosts which are part of the same botnet [5]. Scientific research also relies on DPI for various others applications, such as collecting information about the use of a certain protocol or technology such as video streaming [6], estimating caching benefits [7], or assessing the currently deployed security infrastructure [8].

All these approaches have to deal with the ever-increasing amount of traffic that needs to be processed in current high-speed ISP and enterprise networks. Current traffic volumes and link speeds are especially challenging for sophisticated DPI-based algorithms which perform computationally complex analyses. A number of researchers therefore proposed different approaches for improving the speed of analysis systems [9], [10], distributing the analysis work-load onto multiple cores [9] or machines in a cluster [11], proposed special-purpose hardware [12], or evaluated specific sampling techniques that are targeted at security monitoring processes [13], [14]. While all these approaches show promising results, they force the user to make some trade-offs.

Special purpose hardware yields very high performance, but limits the user to the special type of analysis provided by the hardware. Approaches that parallelize the analysis onto multiple cores or machines require the user to provide the resources (CPU, RAM, bus speed) necessary to analyze all the traffic. In case of a lack of resources, e.g., due to limited budget or due to unforeseen extensive resource consumption by the analysis process, packet loss is inevitable.

Sampling algorithms, on the other hand, can help in scenarios where the available processing power is not sufficient by reducing the amount of traffic that needs to be analyzed. However, many available sampling algorithms need to be configured to select a fixed share of the traffic using a static sampling limit. This sampling limit might be difficult assess, which easily leads to situations where traffic is being discarded, even though the hardware would be able to process a larger share of the traffic. Obviously, this increases the risk that important packets cannot be seen by the monitoring process. A network traffic analysis environment should therefore focus on providing means for fully utilizing the available hardware resources whenever possible, and selecting the appropriate amount of traffic that the hardware setup can handle.

In our work, we present an adaptive load-aware sampling algorithm that is suitable for security monitoring in single-core and multi-core monitoring environments. Our algorithm adapts the number of packets to be sampled according to the currently observed network traffic and the workload patterns of the analysis processes by adapting the sampling limit dynamically. It aims at fully utilizing the available hardware resources, while at the same time trying to sample those packets that are most likely to contain "interesting" content. Fully utilizing the available hardware requires the exploitation of current multi-core hardware. We therefore also focus on an integrated approach that combines our proposed sampling algorithm with current multi-core aware capturing setups, and discuss how to integrate our work into the systems presented in previous work. This work furthermore presents an implementation of the proposed algorithm and evaluates our approach in a real monitoring setup on the 10GE link of a large-scale university network provider.

The remainder of this document is structured as follows. In Section II, we present related work and discuss findings of previous work upon which we build our sampling algorithm. Section III introduces the algorithm and describes the capturing architecture that is used to drive the sampling and the analysis process. It covers the problems of capturing and distributing traffic from the network interfaces onto several application instances. Section IV presents the evaluation of our monitoring setup using real-world online traffic from a 10 Gbit/s Internet uplink. We conclude the paper in Section V.

## II. RELATED WORK

The importance of properly configuring capturing systems in order to get the best performance from the available hardware has been stated and evaluated in previous work [9], [10], [15]. A capturing system consists of several components which all need to perform as fast as possible to avoid packet loss. However, even the best configured and tuned traffic analysis system can still struggle to perform DPI analysis on high-speed links if it is short in computational resources. In this case, packet sampling needs to be employed in order to avoid random packet loss due to resource exhaustion. Our work presents an adaptive load-aware sampling algorithm that helps with fully utilizing the available hardware resources by picking the maximum share of the traffic that the system is able to consume. Therefore, we want to make sure our algorithm can be included into such modern well-configured capturing setups.

In previous work, we analyzed capturing stacks of the FreeBSD and Linux operating systems, including several improvements proposed by researchers, with respect to their capturing performance [10]. Our results concluded that a Linux-based setup that employs the techniques presented by PF_RING [16] and TNAPI [9] provides best capturing performance. Furthermore, these technologies allow for using multiple cores on a single machine with single-threaded applications. Such setups allow to exploit multi-core hardware without requiring a transformation of a monitoring application into a multi-threaded system. Our goal is to develop a sampling algorithm that can be easily included in such kinds of high-performance capturing systems.

The process of picking a subset of the overall traffic for the analysis has been an active research topic in the community Sampling a good subset of packets from the available traffic can be a difficult task. For example in the field of security monitoring: Whenever malicious packets are dropped by the sampling algorithm, an attack or system compromise may go unnoticed as the intrusion detection system does not analyze the non-sampled traffic. Furthermore, previous research shows that sampling can distort anomaly detection metrics [17]. Others found that random packet sampling can have severe impact on the on the detection rates of some analysis algorithms [18].

Different research groups therefore focused on identifying potentially important packets, and aimed at sampling algorithms that select these packets from the overall packet population observed on a link. A general approach to sample traffic for DPI-based inspection systems has been presented and evaluated by different researchers: Several papers have found sampling algorithms that select the first $N$ packets or payload bytes of a flow to result in good detection results when analyzing attack traffic [14], [19], searching for botnet activity [20], or doing computer forensics [21]. Analyzing those first few kilobytes of traffic has also been employed in non-security related work, such as traffic classification [22], analysis of video traffic [23] or SSL traffic [8].

Although research has shown this to be a very promising sampling approach, selecting an appropriate value for the per-flow sampling limit $N$ is difficult: If $N$ is set too low, many interesting packets will be filtered out by the sampling process; whereas if $N$ is too high, system resources will be insufficient to handle all the sampled packets, and packet loss is inevitable. Either way, interesting packets can go undetected which could have been found with a better value for $N$. For attack detection, a major drawback of the approaches is the fact that a malicious attacker can try to send $N$ bytes of legitimate traffic before sending attack packets, thus evading detection [14].

The problem of determining an optimal flow sampling limit $N$ is challenging due to the fact that network traffic tends to undergo dramatic fluctuations, in volume as well as in traffic mixture, on comparably short timescales. These traffic properties result in different traffic features, which in turn change the maximum number of bytes per flow that an intrusion detection or protocol parsing system can handle. In our work, we aim at extending the promising ideas presented in previous work and propose a sampling algorithm that selects the first $N$ bytes, but with $N$ being continuously and dynamically adjusted to reflect the current network and analysis workloads.

Adaptive algorithms have also been studied in previous work. Zhang et al. [13] presented a sampling algorithm that is botnet-aware and tries to sample traffic that likely belongs to a botnet, such as Command and Control traffic. Their approach is application-specific and does not adapt depending on packet consumption rates.

Estan et al. [24] and Barlet-Ros et al. [25] propose sampling algorithms that pick traffic based on available device resources (packet or flow sampling). The goals of both approaches differ from our goals: Their algorithms aim at improving sampling algorithms that generate aggregated statistical information about a link's traffic mix, while we focus on analysing parts of all connections. Our work also differs in the employed techniques: Estan et al. [24] focuses on adapting the sampling rate depending on the used memory resources in fixed pre-configured time bins. A certain pre-defined sampling rate is adopted at the start of each time bin and is lowered throughout the rest of the time interval depending on the consumed memory resources. Barlet-Ros et al. [25] propose a system that adapts the sampling rate based on the needs of the monitoring application. Their system observes traffic features and tries to determine their impact on monitoring application by monitoring its CPU usage. A prediction model is fed with these information and tries to adapts the sampling rate based on the incoming traffic characteristics.

Our approach does not account resource exhaustion, but only monitors the packet consumption rates in order to determine how much traffic the application is able to consume.

## III. SAMPLING ARCHITECTURE AND ALGORITHM

In order to build a sampling algorithm that allows selecting the maximum number of packets the analyzing application(s) is capable to analyze, one has to consider the overall monitoring environment. We introduce our sampling algorithm in Section III-A and discuss in Section III-B the capturing environment this algorithm can be used in.

### A. Adaptive Load-Aware Sampling

In order to determine a good value for the per-flow sampling limit $N$, we need a notion of how many packets the application(s) can handle. Because we do not want to tailor our sampling algorithm to a specific monitoring application, we cannot make any assumptions about the number of packets that an application can consume. Furthermore, even for specific systems such as Snort, these numbers depend on the system configuration, e.g., detection signatures, and the observed traffic features. We therefore need a feedback system that infers the throughput of the analysis system. The fill-level of the buffer between the capturing and the analysis system can serve as an appropriate performance feedback: Whenever the capturing thread inserts packets at a higher rate than the application can consume, the buffer fill-level will rise. This means that the sampling rate must be decreased by reducing $N$ appropriately. On the contrary, if the application is able to consume packets faster than the incoming rate, the buffer will become nearly empty, and the sampling rate can thus be increased.

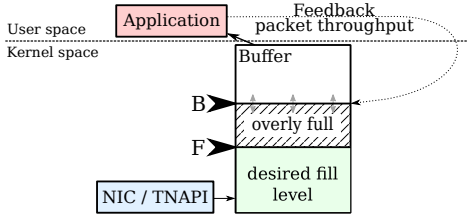

Fig. 1: Buffer fill level feedback mechanism.

When a target buffer fill level $F$ is defined (e.g., on fourth of the overall buffer size), we can measure the deviation $F - B$ of the current fill level $B$ from the target fill level, and use this difference as an indicator for assessing the quality of our current sampling limit. This architecture is illustrated in Figure 1. To decide how to adapt the sampling limit $N$, our algorithm can make use of three indicators:

- Current deviation from the target fill level
- Past deviation from the target fill level
- Estimate for the future development of the fill level

We can use a so-called proportional–integral–derivative (PID) controller as a generic feedback controller to model these indicators. In control theory, PID controllers are very popular and have been shown to be the best form of controller if the system to be regulated cannot be modelled more precisely [26]. This allows us to use this system in a setup where we have to cope with unpredictable input patterns, as well as diverse and unknown application behavior.

A PID controller adjusts a variable according to dynamic development of an input variable, which is in our case the deviation $F - B$. To this end, it does not only consider the input variable, but actually obtains the three correcting input terms $P$, $I$, and $D$. These three terms can be mapped to our three indicators. In our setup, the sampling limit $N$ is the variable that is to be manipulated by the PID controller. It is updated at times $t_0, t_1, ...$, which refer to packet arrivals. The sampling limit at time $t_i$ is calculated as

$$n_{t_i} = const + (k_P \cdot P_{t_{i-1}} + k_I \cdot I_{t_{i-1}} + k_D \cdot D_{t_{i-1}}) \quad (1)$$

The proportional term $P$ reflects the current deviation of the buffer fill level $B$ at time $t_i$ from the desired fill level $F$:

$$P_{t_i} = F - B_{t_i} \quad (2)$$

The integral term $I$ encodes the past deviation from the target fill level:

$$I_{t_i} = \sum_{j=1}^{i} P_{t_j} \cdot (t_j - t_{j-1}) \quad (3)$$

We assume network traffic to be bursty, i.e., quick bursts of packets tend to arrive within very short time intervals. This way, we use $D$ to model extreme changes to the buffer fill level:

$$D_{t_i} = \frac{P_{t_i} - P_{t_{i-1}}}{t_i - t_{i-1}} \quad (4)$$

Each of the $P$, $I$ and $D$ terms is weighted by parameters $k_P, k_I$, and $k_D$, which are used to control the influence of the term. Choosing proper weights $k_{(\cdot)}$ for the individual parameters is an important task for tuning the algorithm.

Networking environments can have substantial short-lived events, such as event-driven packet bursts or the occurrence of large jumbo frames. Such events, though short-lived, can lead to the buffer temporarily filling up very quickly. Such extreme changes in a very short time period bear the potential of sampling limit oscillation and hence unnecessary dropping of packets. In order to mitigate the results of such unforeseeable short-term events, we introduce an additional inertia to the controlling system by applying an exponential moving average (EMA) mechanism. Our new sampling limit does not only depend on the current and past buffer fill level (and its integral and derivatives), but is also influenced by the past sampling limit. We define the final sampling limit $N$ to be constituted from the current PID controller value and the previous sampling limit through

$$N_t = \alpha \cdot n_t + (1 - \alpha) \cdot N_{t-1} \quad (5)$$

for a user-defined parameter $\alpha \in [0,1] \subset \mathbb{R}$, which weights the current controller-calculated sampling limit against the previous sampling limit.

Finding good parameter sets for the algorithm can have heavy influence on the performance of the algorithm. However, as we show in Section IV, a generic parameter set can be found that suits for different monitoring applications under various circumstances. The following heuristics describe the influence of the individual parameters and give hints on how to further tune the parameter sets: The *const* parameter should be set to the desired sampling limit. The $k_P$ parameter should approximately be set such that a full buffer reduces the sampling limit to zero. Increasing the $k_P$ parameter strengthens the impact of current fill level deviations from the desired fill level. A too large $k_P$ parameter can be identified by medium-term oscillations. The $k_I$ parameter can be used to adapt the long-time sampling limit. Increasing its value gives the controller a larger action scope to automatically find a suitable average sampling limit and to cancel out oscillations induced by the $k_P$ parameter. However, it decreases the controller's response time which can be observed by packet loss and may lead to long-term oscillations. Finally, the $k_D$ term increases the controller's response time and acts as counterbalance to the $k_I$ parameter. A too large $k_D$ parameter can be identified by short-time oscillations. The parameter $\alpha$ influences the inertia of the system, with a higher value leading to slower changes to the sampling limit.

### B. Capturing Environment and Sampling Implementation

A capturing setup that optimally utilizes the available processing power must be multi-core aware in order to fully exploit the capabilities of modern commodity hardware. It implies the need for parallelism in capturing and analyzing software. Since one of our algorithm's goals is to adapt the sampling limit in a way that fully utilizes the available processing power, we discuss how to integrate our algorithm in a multi-core-aware capturing and traffic analysis setup.

Based on our previous work where we compared and assessed different capturing systems [10], we decided to build our setup on PF_RING [16] and TNAPI [9]. However, the algorithm itself is not limited to TNAPI but could also be implemented for other approaches such as PFQ [27] or DNA [28]. PF_RING provides an optimized capturing module for Linux, which substitutes the standard AF_PACKET capturing module. TNAPI is a driver improvement that creates a kernel thread for the network interface driver. The threads' only responsibility is to move the captured packets into a buffer that is shared between the kernel and the user space.

In combination with Receive Side Scaling (RSS) techniques, where the network card is able to distribute the incoming traffic across multiple CPU cores, multiple TNAPI threads can be used to perform the capturing [9]. [9] highlights the importance of proper thread scheduling for the involved capturing and analysis threads. Their recommendation is to create one TNAPI thread and one analysis thread for each available core, and to use the same core for the capturing and the analysis thread in order to allow proper use of CPU caches. This recommendation substantially involves the network cards capability to load-balance traffic to the RSS queues. Most cards implement per-flow load-balancing which can result in both directions of a connection to be mapped on different cores. However, many network monitoring applications want to observe both sides of the communication. Software-based load-balancing, which is also supported by PF_RING, is therefore required to achieve a proper biflow-aware load-balancing. This bi-flow mapping can result in several flows being re-mapped onto another core, destroying the cache coherency. In previous work [10], we found that the mapping proposed in [9] is good for light-weight analysis processes, but results in higher packet loss for computational expensive setups. As sampling is important in the latter case, we recommend using different cores for capturing threads and analysis processes.

All these considerations lead to a capturing setup as shown in Figure 2. A number of capturing threads in the network card driver receive packets from the card, and push the packets to multiple user space analysis applications. These threads can belong to the same multi-core aware analysis application, possibly distributed over different process contexts. The capturing threads write the packets into a buffer that is shared between user space and kernel. The application threads read them from these buffers as fast as they can process incoming packets. Our sampling algorithm can be include into this setup as a filtering plugin within the PF_RING sampling architecture and is applied by the capturing threads before the packet is written to the shared buffer between the kernel and the application. A sampling decision is derived on a per-ring fashion: The buffer fill level of each ring is used for the calculation of the sampling limit, resulting in a sampling limit which is adapted for each ring, and hence each application thread. Therefore, thread-specific load characteristics are considered by the sampling process.

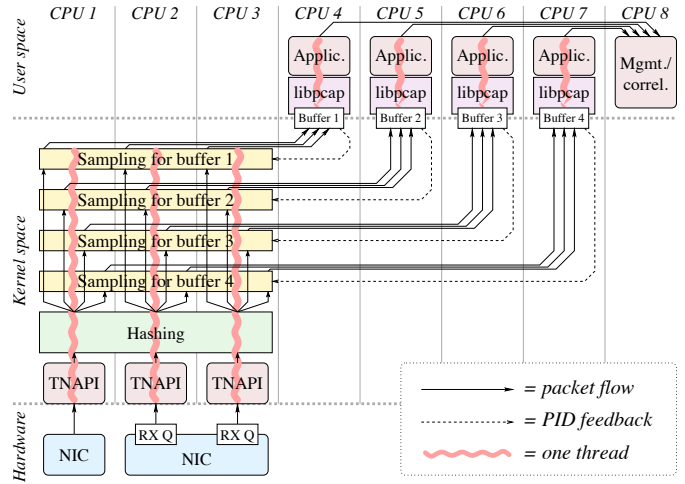

Fig. 2: Data flow for an example setup.

## IV. EVALUATION

Since our algorithm is influenced by packet arrival times and application's processing time per packet, an evaluation should be performed on a real productive system on real network traffic. This is especially important since application processing times depend on the used hardware, monitoring application, and observed traffic features. We therefore start our evaluation with a description of our hardware setup and the network where we were able to deploy our vantage point in Section IV-A. The validation of our algorithm in live experiments with different monitoring applications on our 10GE link is described in Section IV-B and Section IV-C with two setups.

### A. Evaluation Setup

A live capturing setup for our evaluation was deployed in the *Munich Scientific Research Network* (Münchner Wissenschaftsnetz, MWN) in Munich, Germany. The research network interconnects three major universities and several affiliated research institutions in the area in and around Munich. Furthermore, the network includes several student dorms that provide housing for the students enrolled in the universities in Munich. Finally, the network hosts a large super-computing cluster that is used by researchers from Munich and other research facilities around the world. In total, the network hosts about 100,000 devices which are used by approximately 120,000 users. It is operated by the *Leibniz Supercomputing Center* (Leibniz-Rechenzentrum, LRZ) and provides Internet access for all its users via a 10 GBit/s link to its upstream provider the *German research network (DFN)*. Our vantage point was deployed on the border gateway between the MWN and its upstream service provider and is therefore able to observe traffic exchanged between the MWN and the Internet.

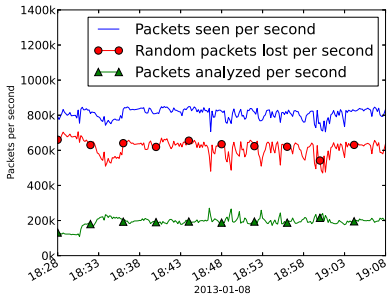

Fig. 3: Snort – Single Instance – Without sampling – *botcc*

TABLE I: Sampling parameter set for Snort

| Parameter | Value |
|---|---|
| $k_P$ | 3333335 |
| $k_I$ | 0.00093 |
| $k_D$ | 1500 |
| *const* | 1.1 MB |
| $\alpha$ | 0.2 |
| buffer size | 268 MB |

Our monitoring setup was built around standard off-the-shelf PC hardware, operated by a Linux-based operating system. It was bought in 2009 and features a 3.2GHz Intel Core i7 processor with four cores and hyperthreading enabled. Kernel and user space share a total of 12 GB of RAM that have been built into the system. An Intel 10GE network card based on the 82598EB chipset is used in conjunction with a TNAPI driver [9]. Three virtual CPUs are allocated to the TNAPI driver to perform capturing, traffic distribution onto the analyzing processes, and sampling. The other cores are able to run instances of the analyzing application. We implemented our algorithm as a PF_RING filtering plugin[1], executing in the softIRQ context of the TNAPI threads.

In the following sections, we evaluate our algorithm in live setups. We use Snort in different configurations to perform extensive live packet analysis on our 10GE link. Snort [1] is a well-known security monitoring toolkit for signature-based traffic analysis. It is known for its pattern matching capabilities, and has been used for a number of publications in the context of network security monitoring. Its traffic analysis engine is driven by rule sets that define payload patterns of interest that point to malicious activity in the network. Searching for those patterns in traffic payload is known to be a computationally complex task and can be challenging on high-speed and high-bandwidth networks. The complexity of the analysis that Snort performs depends on the configuration and the rule set, which defines the patterns that Snort searches in the analyzed traffic. Hence, we can use Snort to evaluate our algorithm in different load scenarios with simple and computationally complex setups.

We show the results for two different configurations with different complexity. These configurations will be called *botcc* and *fullset* in the remainder of this work. Both were obtained on Jan. 08 2013 from emergingthreats.net [29]. The lightweight *botcc* configuration is the free "ETopen" rule set that contains rules for botnet command and control traffic detection. It includes 146 rules with IP addresses of known command and control servers of different botnets. Snort does not have to perform pattern matching with this configuration but must only check IP addresses of the observed packets against the list of command and control servers. The *fullset* consists of the complete set of 11,748 rules, which contain a lot of patterns to match against the observed packets. This configuration is used as a setup to demonstrate the applicability of our algorithm for complex traffic analysis tasks that put high load on the traffic processing engine.

### B. *Simple analysis with lightweight rule set* botcc

Figure 3 compares the number of packets on the link to the number of packets that could be analyzed by Snort with the *botcc* configuration. The figure shows the number of packets per second (pps) that have been captured on the link with the blue line. During the approximately 30 minutes monitoring period, this number of packets varies between 705k and 850k pps, which corresponds to the maximum number of packets per seconds that can be delivered by the border gateway router. On average the incoming packet rate was around 810k pps (mean) or 820k pps (median) during the observation period. Only a single Snort instance was used to analyze the complete traffic on the link. The number of analyzed packets is by far lower as the incoming rate and changes over time depending on the features of the incoming traffic (e.g. number of packets, number of new connections, etc). It varies between about 270k pps (max) and 108k pps (min) at an average of 195k packets per second. Snort's packet processing rate results in a median packet loss rate of about 630k pps. As previous research discussed in detail, e.g. [20], [19], [10], this kind of random packet loss has the potential of losing interesting packets.

We started another monitoring run with the same Snort configuration, but this time we enabled the sampling algorithm. Table I shows the PID controller parameters used throughout our experiments. They were determined in experiments using the heuristics of Section III-A. These parameters were used for all experiments of the paper, including the multi-instance setups and setups with different rule sets presented later on. Our choice of parameters may not be optimal but the results reveal that our solution performs well in different scenarios even with a non-optimal choice of parameters.

Figure 4 plots the packet rate statistics for traffic analysis in this setup. Incoming packet rates were similar compared to the previous run: 860k pps were observed during peak times, minimum packet rates were around 600k pps with a median of 811k pps. The number of packets sampled by our algorithm matches the number of packets consumed by Snort. No random packet loss due to full buffers occurred, and all packets picked by the sampling algorithm could be analyzed by Snort. This reveals that our sampling algorithm picks the appropriate sampling limit. A closer look at the number of Snort's packet consumption shows a mean number of around 115k pps with a maximum of 213k pps and a minimum of 56k pps. The median packet consumption rate is 105k pps. One can see that the average packet processing rate in our sampling setup is lower than the processing rate without sampling. The reason for this can be found when taking a closer look at the analysis that Snort performs, which results in non-uniform packet processing. Instead, the per-packet processing time depends on the characteristics of the incoming traffic. Our sampling algorithm picks packets from the beginning of the connections, which forces Snort to observe all connections on the network. Random packet loss tends to oversample connections with much traffic and tends to miss shorter flows [30]. As Snort keeps per-connection state, this random packet loss decreases the number of internal state Snort needs to manage and thus influences the packet processing time. Those effects are in line with the findings of previous work which analyzed the per-packet processing rates of Snort [31]. The authors find that the packet processing times for packets at the beginning are higher than those packets at later points in a session. Hence, the lower number of processed packets is an indicator that shows our sampling algorithm works as expected. Similar effects will be observable in the multi-instance setup and for the *fullset*.

---

[1]https://github.com/diekmann/cctrack

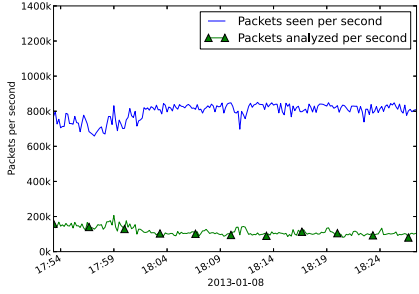

Fig. 4: Snort – Single Instance – Sampling – *botcc*

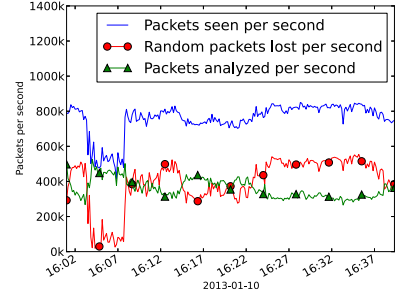

Fig. 6: Snort – 4 instances – Without sampling – *botcc*

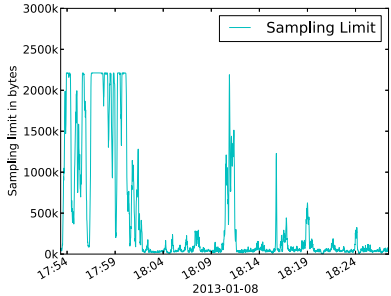

Fig. 5: Sampling limit development – Single Instance – *botcc*

We can observe the behavior of the PID controller when inspecting the development of the sampling limit during this particular monitoring period. Our sampling limit always counts the complete layer four payload including header (TCP/UDP). The first packet that exceeds the sampling limit is passed entirely to the application; only subsequent packets are dropped. Therefore, the first packet of a flow is always sampled. Figure 5 plots the sampling limit development throughout the complete monitoring run. The sampling limit starts with our *const* parameter of 1.1 MB sampled traffic per flow, and adapts according to the incoming packet rate and the processing rate of Snort. As the sampling limit is updated on the arrival of a sampled packet, a large number of sampling limits changes can be observed during the monitoring interval. Our kernel module provides average statistics on its sampling limit: Every second a mean value of the sampling limits of the last second is generated.

One can observe heavy changes in the first 10 minutes of the sampling interval, which correlates with the incoming packets rates seen in Figure 4. Incoming packet rates in this time interval vary highly between 800k pps and 650k pps. The sampling limit adapts to these packet rates: it increases as capturing buffers empty during lower incoming packet rates, and decreases as buffers fill up due to increased incoming packet rates. During this time, the sampling limit reaches a limit set by the implementation and its configuration. As our algorithm is implemented as a kernel-level filter, we cannot use floating-point arithmetic [32] but need to calculate sampling limit updates using integer arithmetic. The maximum sampling limit depends on the parameter set, e.g. buffer size, *const*, $k_{(\cdot)}$, $\alpha$. For our configuration, the maximum sampling limit is reached at around 22 MB.

As packet rates grow steadier towards the maximum number of packets the system can consume, changes to the sampling limit become smaller and the sampling limit settles down at much smaller values. However, we can observe increases in the sampling limit as soon as the number of packets on the link decreases. Hence, the system tries to maximize the number of analyzed packets, which is the improvement we desired over the traditional Time-Machine style sampling.

In order to study our algorithm in modern multi-core-aware environments, we set up a test run with four instances of the same Snort configuration (*botcc*) running in parallel. At first, we use the configuration with our sampling disabled, and report the performance of the parallel running Snort instances in Figure 6. The incoming packet rate, blue line, starts at around 800k pps and remains in this region throughout most of the time. However, at the beginning of the monitoring interval, starting at about 16:05, there is an approximately 5 minute time interval where the incoming packet rate drops to around 500k pps. The four Snort instances consume, as expected, a significant higher packet rate, averaging at 360k pps (mean) or 350k pps (median). However, more than half of the packets (396k pps mean / 433k pps median) are dropped due to full buffers. Again, this random packet loss reduces the number of flows that can be observed by the analyzers.

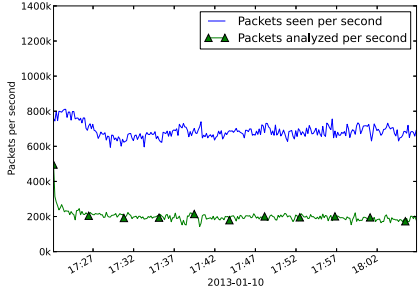

Fig. 7: Snort – with sampling – 4 instances – *botcc*

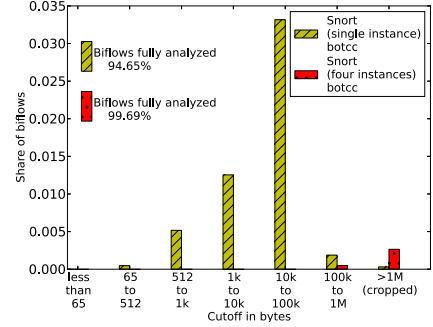

Fig. 9: Biflow cut off – *botcc*

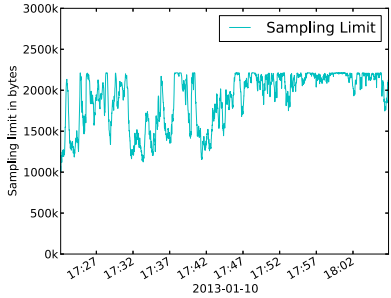

Fig. 8: Sampling limit development – 4 instances – *botcc*

When adding our sampling algorithm to the setup, as shown in Figure 7, we can see some changes to the processing rate. At first, the number of incoming packets is smaller than in the run before. It averages at a mean of about 686k pps (median: 678k pps), and varies between 840,000 and 594,000 pps. All Snort instances are able to consume the complete set of sampled packets, at rates that vary between 496k pps and 132k pps with mean and median at about 198k pps. No random packet loss occurred. A closer look at the development of the sampling limit, shown in Figure 8, reveals an increased sampling limit compared to the first monitoring run with a single instance. This is due to the lower incoming packet rate, and the fast packet consumption by Snort. All obtained average sampling limits in our monitoring interval are higher than 10 MB.

As our plotted sampling limit is a per-second average, the applied per packet sampling limit may vary from the average in certain cases. We therefore cross-check the influence of the sampling limit by examining the biflow cutoff during the setups. A biflow cutoff is the point in the biflow, where the sampling algorithm decides to cut off the connection and stops sampling more traffic from this biflow. This cutoff depends on the current sampling limit when a new packet is observed for a biflow.

Figure 9 shows the sampling limit cutoff for the previous scenarios for all observed biflows. The single instance setup observed 27.8 million biflows throughout its monitoring period. Out of these, 94.65% were fully analyzed, which means that no sampling cutoff applied for these biflows. The remaining share of the flows where cut off as shown in the figure. The packet for which the cutoff occurs is always sampled, thus, the first packet(s) of each flow are always passed to the application. The class $>1M$ shows those flows that have not been fully sampled but have had a cutoff that is larger than 1 MB. In the single instance monitoring run, only very few biflows where cut off at 65 to 512 bytes. Most of the biflows that where not fully sampled had a cutoff between 10KB and 100KB. However, even this class represents less than 0.1% of all flows. The four instance Snort monitoring scenario with its larger sampling limit had 99.69% of all flows completely sampled.

### C. Complex traffic analysis with rule set fullset

In order to determine the results of our algorithm in higher-load scenarios, we performed experiments with the *fullset* rule set. As this rule set includes many more rules than the previous one, we would expect a lower number of packets analyzed by Snort. The results in Figure 10a confirm this assumption: A median incoming packet rate of 756k pps translates into a median loss rate of 707k pps.

Figures do not increase significantly when four instances are run instead of a single one, as shown in Figure 10b. This time, an incoming packet rate of 780k pps results in a median loss rate of 659k pps. While adding more analysing instances increases the analyzed packet rate from 114k pps to 179k pps, the majority of the traffic is still randomly lost.

When we observe the incoming packet rates for the single and multi instance setups with *fullset* with our sampling enabled, we make similar observations as before. Figure 10d shows the figure for the single and multi instance setups. Both have similar incoming packet rates with a median at 781k pps (single instance) and 734k pps (multiple instances), and both setups did not experience any data loss due to full buffers. All packets that have been taken by the sampling algorithm could be analyzed by the Snort instances. The multi instance setup analyzed a median packet rate of 84,000 pps while the single setup only analyzed 58,000 pps.

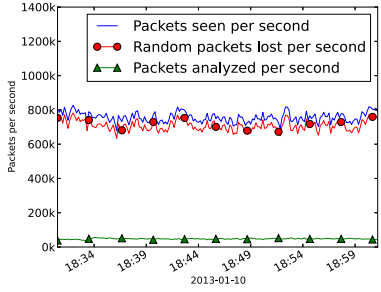

Fig. 10a: Snort – Single Instance – Without Sampling – *fullset*

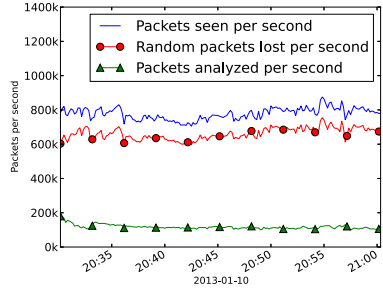

Fig. 10b: Snort – Multiple instances – Without Sampling – *fullset*

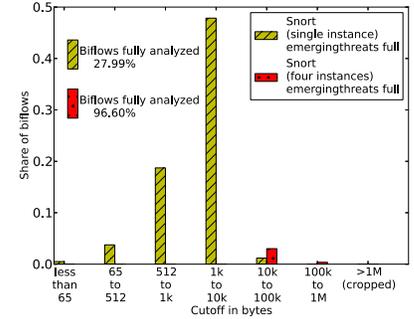

Fig. 10c: Biflow cutoff – *fullset*

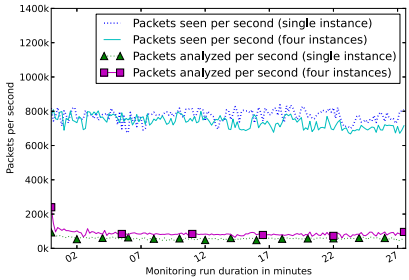

Fig. 10d: Snort – With Sampling – *fullset*

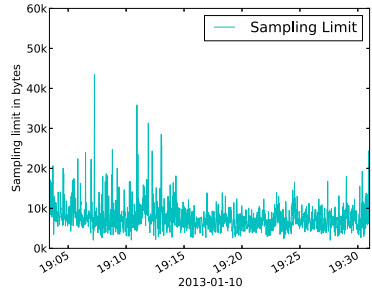

Fig. 10e: Snort – Sampling Limit – Single Instance – *fullset*

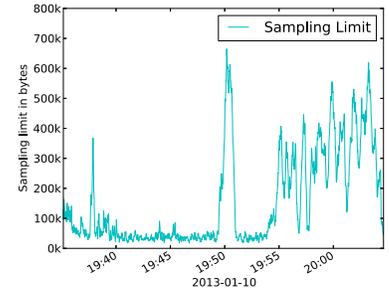

Fig. 10f: Snort – Sampling Limit – Multple Instance – *fullset*

A closer look at the sampling limit development reveals bigger differences than the look at the packet consumption rate. Figure 10e and Figure 10f compare the sampling limit for both setups. While the sampling limit for the single Snort setup can be measured in the region of 10Kb, the four instance setup finds itself with a higher average limit.

The same observation can be made when analyzing the biflow cutoff, which is shown in Figure 10c for both setups. The single instance setup, where only one core was used to analyze the complete traffic, was only able to analyzed around 28% of the observed 20,633,534 million biflows completely, and had to cut off the rest of the biflows. Most of the biflows were analyzed to an extend between 1KB to 10KB of their length. The multi instance setup was capable to fully analyze 96.6% of its 20,134,556 million biflows in its monitoring interval. All remaining biflows were cut off somewhere after 100KB with no biflow being cut off with less than 100KB observed.

## V. CONCLUSION AND FUTURE WORK

This paper presented an adaptive load-aware sampling algorithm for high speed networks. It is based on a well-known sampling mechanism that previous work found to be useful for many applications. Our proposed algorithm overcomes limitations of previous work by introducing a dynamic sampling limit. This sampling limit is automatically adapted to match run-time events such as changes in the incoming packet rate or packet consumption rates of the monitoring application. It is chosen such that both the monitoring application's utilization of processing power is maximized and the random packet loss is minimized.

We evaluated our algorithm in live traffic capturing setups and found that traffic feature changes make choosing a static sampling limit required by previous algorithms very difficult. Short-term changes in traffic features can influence the processing rate of monitoring applications such that that they are able to consume more or less traffic than in the previous time interval. Our dynamic sampling algorithm adapts the sampling limit to those short-term events with the result that the available computing resources are fully utilized. It is capable of being used in current multi-core aware traffic setups which run multiple instances of monitoring applications, each on a separate core. These capabilities allow the inclusion of our algorithm into traffic analysis setups that exploit the features of current multi-core hardware.

Our work presented heuristics for determining a good parameter set for our algorithm, and our evaluation showed that a generic parameter set yields good performance in different scenarios. Providing the user with a mechanism that automatically finds optimal parameters for a given capturing setup could further improve the ease of use and performance of our algorithm. Furthermore, our algorithm targets all monitoring applications that could benefit from sampling the beginning of connections. Its performance was therefore evaluated with regard to analyzed packets, packet drops, and the development of the sampling limit. This evaluation can be extended by a closer look at the influences of our sampling algorithm on specific applications, such as detection rates in security monitoring tools.

## VI. Acknowledgments

This work has been partially funded by the German Federal Ministry of Education and Research (BMBF) within the projects SASER and Peeroskop.

## References

[1] Martin Roesch. Snort – Lightweight Intrusion Detection for Networks. In *Proceedings of the 13th USENIX Conference on System Administration (LISA '99)*, Seattle, Washington, November 1999.

[2] Vern Paxson. Bro: A System for Detecting Network Intruders in Real-Time. *Computer networks*, 31(23-24):2435–2463, 1999.

[3] Guofei Gu, Philip Porras, Vinod Yegneswaran, Martin Fong, and Wenke Lee. BotHunter: detecting malware infection through IDS-driven dialog correlation. In *Proceedings of 16th USENIX Security Symposium*, page 12, 2007.

[4] Guofei Gu, Junjie Zhang, and Wenke Lee. BotSniffer: Detecting Botnet Command and Control Channels in Network Traffic. In *Proceedings of the 15th Annual Network and Distributed System Security Symposium (NDSS'08)*, February 2008.

[5] Tiang-Fang Yen and Michael K Reiter. Traffic aggregation for malware detection. In *Detection of Intrusions and Malware, and Vulnerability Assessment*, pages 207–227, Paris, France, July 2008. Springer.

[6] Alessandro Finamore, Marco Mellina, Maurizio M. Munafo, Ruben Torres, and Sanjay G. Rao. YouTube Everywhere: Impact of Device and Infrastructure Synergies on User Experience. In *Proceedings of the 11th Annual Conference on Internet Measurement (IMC '11)*, Berlin, Germany, November 2011.

[7] Bernhard Ager, Fabian Schneider, Juhoon Kim, and Anja Feldmann. Revisiting Cacheability in Times of User Generated Content. In *INFOCOM IEEE Conference on Computer Communications Workshops*, 2010.

[8] Ralph Holz, Lothar Braun, Nils Kammenhuber, and Georg Carle. The SSL Landscape: A Thorough Analysis of the X.509 PKI Using Active and Passive Measurements. In *Proceedings of the 11th Annual Conference on Internet Measurement (IMC '11)*, Berlin, Germany, November 2011. ACM.

[9] Francesco Fusco and Luca Deri. High Speed Network Traffic Analysis with Commodity Multi-Core Systems. In *Proceedings of the 10th Annual Conference on Internet Measurement (IMC '10)*, Melbourne, Australia, November 2010.

[10] Lothar Braun, Alexander Didebulidze, Nils Kammenhuber, and Georg Carle. Comparing and Improving Current Packet Capturing Solutions based on Commodity Hardware. In *Proceedings of the 10th Annual Conference on Internet Measurement (IMC '10)*, Melbourne, Australia, November 2010.

[11] Matthias Vallentin, Robin Sommer, Jason Lee, Craig Leres, Vern Paxson, and Brian Tierney. The NIDS cluster: scalable, stateful network intrusion detection on commodity hardware. In *Proceedings of the 10th International Conference on Recent Advances in Intrusion Detection (RAID '07)*, Gold Coast, Australia, September 2007.

[12] Nicholas Weaver, Vern Paxson, and Jose M. Gonzalez. The Shunt: An FPGA-Based Accelerator for Network Intrusion Prevention. In *Proceedings of the 2007 ACM/SIGDA 15th International Symposium on Field Programmable Gate Arrays (FPGA '07)*, Monterey, CA, February 2007. ACM.

[13] Junjie Zhang, Xiapu Luo, Roberto Perdisci, Guofei Gu, Wenke Lee, and Nick Feamster. Boosting the Scalability of Botnet Detection using Adaptive Traffic Sampling. In *Proceedings of the 6th ACM Symposium on Information, Computer and Communications Security (ASIACCS '11)*. ACM Request Permissions, March 2011.

[14] Lothar Braun, Gerhard Münz, and Georg Carle. Packet Sampling for Worm and Botnet Detection in TCP Connections. In *12th IEEE/IFIP Symposium on Network Operations and Management Symposium (NOMS 2010)*, Osaka, Japan, April 2010.

[15] Fabian Schneider, Jörg Wallerich, and Anja Feldmann. Packet Capture in 10-Gigabit Ethernet Environments using Contemporary Commodity Hardware. In *Proceedings of the 8th International Conference on Passive and Active Network Measurement (PAM '07)*. Springer-Verlag, April 2007.

[16] Luca Deri. Improving Passive Packet Capture: Beyond Device Polling. In *Proceedings of the 4th International System Administration and Network Engineering Conference (SANE '04)*, Amsterdam, The Netherlands, September 2004.

[17] Daniela Brauckhoff, Bernhard Tellenbach, Arno Wagner, Martin May, and Anukool Lakhina. Impact of Packet Sampling on Anomaly Detection Metrics. In *Proceedings of the 6th ACM SIGCOMM Conference on Internet Measurement (IMC) 2006*, pages 159–164. ACM, 2006.

[18] Jianning Mai, Ashwin Sridharan, Chen-Nee Chuah, Hui Zang, and Tao Ye. Impact of packet sampling on portscan detection. *Journal on Selected Areas of Communication (IEEE)*, 24(12):2285–2298, 2006.

[19] Gregor Maier, Robin Sommer, Holger Dreger, Anja Feldmann, Vern Paxson, and Fabian Schneider. Enriching Network Security Analysis with Time Travel. *Proceedings of the ACM SIGCOMM 2008 Conference on Data Communication (SIGCOMM '08)*, August 2008.

[20] Konrad Rieck, Guido Schwenk, Tobias Limmer, Thorsten Holz, and Pavel Laskov. Botzilla: Detecting the "Phoning Home" of Malicious Software. In *Proceedings of the 2010 ACM Symposium on Applied Computing (SAC '10)*, Sierre, Switzerland, March 2010.

[21] Stefan Kornexl, Vern Paxson, Holger Dreger, Anja Feldmann, and Robin Sommer. Building a Time Machine for Efficient Recording and Retrieval of High-Volume Network Traffic. *Proceedings of the 5th ACM SIGCOMM Conference on Internet Measurement (IMC '05)*, pages 23–23, October 2005.

[22] Giuseppe Aceto, Alberto Dainotti, Walter De Donato, and Antonio Pescapé. PortLoad: taking the best of two worlds in traffic classification. In *IEEE Conference on Computer Communications (INFOCOM), Workshops, 2010*, San Diego, CA, March 2010. Ieee.

[23] Lothar Braun, Alexander Klein, Georg Carle, Helmut Reiser, and Jochen Eisl. Analyzing Caching Benefits for YouTube Traffic in Edge Networks - A Measurement-Based Evaluation . In *13th IEEE/IFIP Symposium on Network Operations and Management Symposium (NOMS 2012)*, pages 1–8, Maui, Hawaii, April 2012.

[24] Cristian Estan, Ken Keys, David Moore, and George Varghese. Building a better NetFlow. *SIGCOMM Comput. Commun. Rev.*, 34(4):245–256, August 2004.

[25] Pere Barlet-Ros, Gianluca Iannaccone, Josep Sanjuàs-Cuxart, Diego Amores-López, and Josep Solé-Pareta. Load shedding in network monitoring applications. In *2007 USENIX Annual Technical Conference on Proceedings of the USENIX Annual Technical Conference*, ATC'07, pages 5:1–5:14, Berkeley, CA, USA, 2007. USENIX Association.

[26] Stuart Bennett. *A history of control engineering, 1930–1955*. Institute of Electrical Engineers (I.E.E.), 1993.

[27] Nicola Bonelli, Andrea Di Pietro, Stefano Giordano, and Gregorio Procissi. On Multi-Gigabit Packet Capturing with Multi-Core Commodity Hardware. In *13th International Conference on Passive and Active Measurement (PAM 2012)*, Vienna, Austria, March 2012.

[28] Direct NIC Access Website. http://www.ntop.org/products/pf_ring/dna/. visited: Jan. 2013.

[29] *Emerging Threats*. http://www.emergingthreats.net/. visited: Jan. 2013.

[30] Shijin Kong, Tao He, Xiaoxin Shao, and Xing Li. Time-Out bloom Filter: A New Sampling Method for Recording More Flows. *Proceedings of the International Conference on Information Networking (ICOIN '06)*, January 2006.

[31] Javier Verdú, Jorge Garcí, Mario Nemirovsky, and Mateo Valero. Architectural Impact of Stateful Networking Applications. In *Proceedings of the 2005 ACM symposium on Architecture for Networking and communications systems*, ANCS '05, pages 11–18, New York, NY, USA, 2005. ACM.

[32] Rusty Russell. Unreliable Guide To Hacking The Linux Kernel. http://kernel.org/doc/htmldocs/kernel-hacking.html, 2005. visited: Jan. 2013.